\documentclass[prl,twoside,onecolumn,notitlepage,superscriptaddress]{revtex4-1}
\usepackage{graphicx}
\usepackage{realhats}
\usepackage[utf8]{inputenc}
    \begin{document}
\title{Percolation and Dissolution of Borromean Networks}
\author{$\hat[cowboy]{\text{D}}$onald G. Ferschweiler}
\affiliation{Department of Physics and Astronomy, California State University, Long Beach}
\author{Ryan Blair} 
\affiliation{Department of Mathematics and Statistics, California State University, Long Beach}
\author{Alexander R. Klotz\footnote{alex.klotz@csulb.edu}}
\affiliation{Department of Physics and Astronomy, California State University, Long Beach}

\begin{abstract}
Inspired by experiments on topologically linked DNA networks, we consider the connectivity of Borromean networks, in which no two rings share a pairwise-link, but groups of three rings form inseparable triplets. Specifically, we focus on square lattices at which each node is embedded a loop which forms a Borromean link with pairs of its nearest neighbors. By mapping the Borromean link network onto a lattice representation, we investigate the percolation threshold of these networks, (the fraction of occupied nodes required for a giant component), as well as the dissolution properties: the spectrum of topological links that would be released if the network were dissolved to varying degrees. We find that the percolation threshold of the Borromean square lattice occurs when approximately 60.75\% of nodes are occupied, slightly higher than the 59.27\% typical of a square lattice. Compared to the dissolution of Hopf-linked networks, a dissolved Borromean network will yield more isolated loops, and fewer isolated triplets per single loop. Our simulation results may be used to predict experiments from Borromean structures produced by synthetic chemistry.
\end{abstract}

\maketitle

\section{Introduction}

Borromean rings consist of three mutually entangled loops for which no two loops share a binary entanglement. They were originally the heraldic symbol of the House of Borromeo and feature prominently in centuries of art, can be constructed from organometallic synthesis or DNA origami \cite{cantrill2005nanoscale}, and serve as an analogy for certain bound states in nuclear physics \cite{bertulani2007geometry}. The three-loop rings are not the only such topological structure, and larger components with Borromean connectivity can be constructed. Large, regular, planar structures are referred to as \textbf{Borromean networks}, or Borromean lattices or chainmail, which are depicted with square lattice symmetry in Figure 1. This extends the existing concept of the Olympic gel, a network of Hopf-linked molecular rings \cite{igram2016resolving}, into a system with only Borromean connectivity.  Borromean lattices are not Brunnian in that the removal of a single component may remove several other but will not dissolve the network entirely.  Physical Borromean lattices have been created by synthetic chemists \cite{thorp2015infinite, hardie2016self, catalano2021open}, and the crystallographic symmetry of Borromean lattices has been discussed \cite{o2020crystallographic,o2021borromean}.

\begin{figure}
    \centering
    \includegraphics[width=0.8\textwidth]{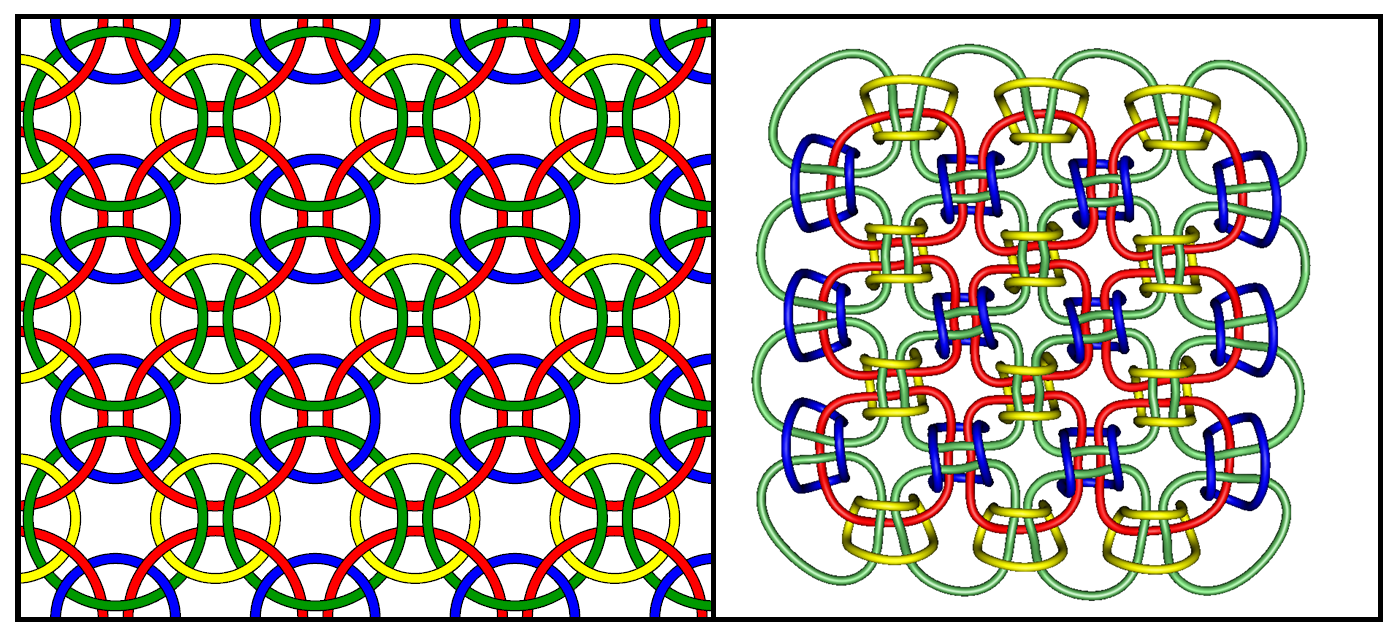}
    \caption{Left: Stylistic rendering of a Borromean chainmail square lattice by Luc Devroye \cite{luc}. No two rings share a direct topological link, but no ring can be removed from the network. Right: Rendering of such a network in three dimensions, visualized and annealed using KnotPlot.}
    \label{fig:chainmail}
\end{figure}

\textbf{Kinetoplasts} are complex topologically linked networks of  DNA molecules found in the mitochondria of trypanosome parasites. Each kinetoplasts consists of about  5000 linked DNA ``minicircles'' which decode the RNA produced by a few dozen linked ``maxicircles'' \cite{shapiro1995structure}. Each minicircle in a kinetoplast is Hopf-linked (the simplest linking between two circles, as in the Olympic emblem) to its spatial neighbors. The average valence, the number of Hopf links that each minicircle shares with its neighbors, is estimated at three, consistent with a honeycomb lattice topology \cite{chen1995topology, ibrahim2018estimating}. These estimates are based on experiments that dissolve the network using enzymes that break individual minicircles, and measure the relative frequency of different topological structures by gel electrophoresis. The network topology is determined by comparing the relative frequency of different components to the predictions from graph theory models. The graph theory models treat each minicircle as a node and each Hopf link between minicircles as an edge between two nodes. This allows predictions, for example, of the probability that a single minicircle is released from the network as a function of the number of minicircles that are dissolved. It is not expected that kinetoplasts have Borromean connectivity, and no such experiments have been performed for Borromean materials. Simulations of dense circle packing \cite{rodriguez2013percolation, diao2012effects} or ring polymer interlocking \cite{michieletto2015kinetoplast,d2017linking}, often applied to percolation phenomena in kinetoplast DNA, typically compute only the Gauss linking number which detects pairwise interactions, but cannot detect Borromean connectivity. Here, we consider the outcome of comparable dissolution experiments with Borromean networks.

A lattice with regular geometry may be regarded as full if each geometric site on the lattice is occupied by a node in a network or graph that shares edges with its geometric neighbors. A full lattice has a single component that spans the entire length scale of the network, known as the giant connected component. As nodes or edges are removed from the network, smaller clusters will be isolated when none of their nodes share edges with the giant component. If a sufficient fraction of edges or nodes is removed, the largest component in the system will no longer span the length scale of the system, and the giant connected component will cease to exist. The minimum fraction of filled nodes or edges in a graph that allows a giant connected component to exist is known as the \textbf{percolation threshold} \cite{essam1980percolation} and depends on the topology of the graph (for example, a square lattice requires 59.27\% of its sites occupied to reach percolation \cite{ziff1992spanning}). Systems at the percolation threshold have universal properties (such as a cluster size distribution exponent and a fractal dimension) that depends only on space dimensionality but not graph topology \cite{mertens2017universal}.


In this work, we explore the connectivity of Borromean lattices through a computational graph theory model. We apply this model to square lattice Borromean networks and analyze the cluster size distribution of Borromean networks as individual loops are removed from the system. We compare these results to the dissolution of a traditionally connected graph, which serves as an analogy to Hopf-linked topological lattices. We investigate the percolation threshold of Borromean lattices, comparing it to the standard square lattice behavior expected from a Hopf-linked network and the universal properties of two-dimensional percolation. We make predictions of the distribution of the smallest clusters that would be isolated from the network: single loops and three-component Borromean rings, and consider whether an experiment could determine whether a topological network is Hopf-linked or Borromean based on a dissolution reaction.  Although it is not expected that kinetoplasts or HK97 viral capsids (another topologically linked biomolecular structure \cite{wikoff2000topologically}) have Borromean connectivity, such analysis can provide insight about whether a given network is Borromean or Hopf-linked.

\section{Mapping and Computation}


We represent topologically linked networks as a square lattices with each node representing an individual loop, similar to Figure 1. With ``standard'' connectivity rules, two occupied sites on a lattice are considered adjacent (sharing an edge in a graph) if they are orthogonally next to each other. This can be taken as analogous to two loops sharing a Hopf link. Unlike a Hopf-linked network with standard pairwise connectivity, Borromean lattice contains no pairwise adjacency, at least three loops are required to establish connectivity. Instead, we define Borromean connectivity, where groups of three occupied sites are considered mutually adjacent if they form the smallest possible right triangle on the lattice, such that the three loops represented by the appropriate space curves at these sites would form Borromean rings. In a graph representation of the lattice, each triangle forms a cluster with three nodes and three edges, while nodes that are not part of a triangle share no edges with other nodes. Connected nodes can be part of more than one triangle, as clusters in the network can contain more than three rings.

\begin{figure}
    \centering
    \includegraphics[width=0.8\textwidth]{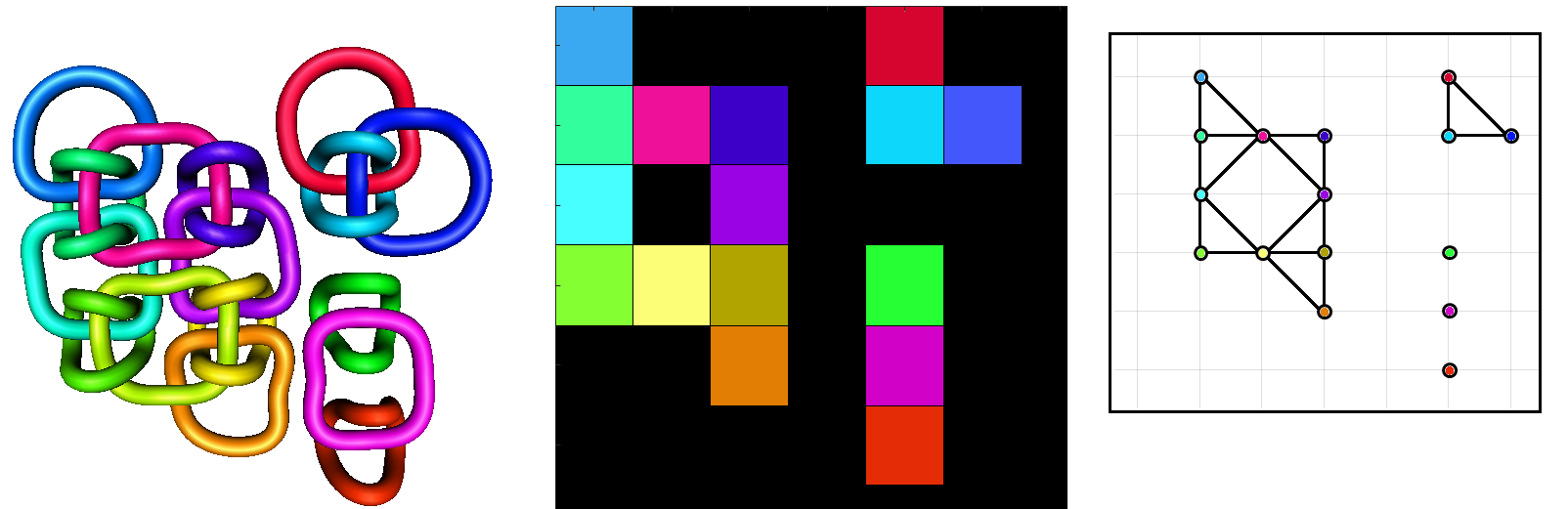}
    \caption{Mapping between the Borromean lattice and our graph theory representation. Left: A Borromean network with a large cluster of ten rings, a triplet of traditional Borromean rings, and three isolated rings. Middle: Representation of the Borromean network on a square lattice where black sites are unoccupied and occupied sites are color-coded with the ring they represent. Right: Graph representation in which each node corresponds to an occupied lattice site, and three nodes share edges if their corresponding lattice sites form a right triangle with unit-length sides, meaning their corresponding rings are Borromeanly linked.}
    \label{fig:mapping}
\end{figure}

Figure 2 shows the mapping between a network of Borromeanly-linked rings, the square lattice representation, and the graph representation. The ring network contains a 10-ring component, a 3-ring Borromean component, and three unlinked rings in spatial proximity. On the lattice, each occupied site in the large component can form a right triangle with at least two of its neighbors, as can each component in the Borromean triplet, but the three singlets cannot. In the graph representation, edges are not shared between the unlinked components. 

The mapping between ring network, lattice, and graph allows an efficient simulation of incomplete Borromean networks. Networks are initialized as random binary matrices of width $L$ representing a square lattice with unoccupied and occupied sites, and can be converted to the 3D representation by substituting a space curve at each lattice site, or converted to the graph representation by calculating which occupied sites form triangles.


On a square lattice, each node has eight neighbors and may form twelve distinct nearest-neighbor triangles with them. Our simplest algorithm scanned each node of the lattice and checked whether the neighboring nodes required to form each of the twelve triangles were occupied. If they were, three bits were flipped in an adjacency matrix, indicating that the three nodes are involved in a Borromean connection, and an additional three were flipped to keep the matrix symmetric under transposition. 

A cluster size distribution for each lattice instance is determined from the adjacency matrix using a reverse Cuthill-McKee algorithm \cite{liu1976comparative, reverse}. To distinguish between unoccupied sites and occupied but isolated sites (which would not register on an adjacency matrix), the rows and columns corresponding to empty sites were removed from the adjacency matrix before cluster size calculation. 

For traditional square lattice connectivity, representing Hopf-link connections, a graphics processing algorithm that uses a flood-fill technique exists to quickly determine connected components (``bwconncomp'' in MATLAB), which is asymptotically faster than the adjacency matrix cluster calculation used for the Borromean lattices. In practice, the Borromean calculations can be sped up by first finding the components with the traditional Hopf connectivity rule, then creating a sub-matrix of each square component and applying the Borromean clustering algorithm to it. Typically a lattice with $L^2$ sites requires an adjacency matrix with $L^4$ total sites. At percolation, the largest square lattice cluster size grows as approximately $0.45\times L^{1.9}$ meaning the Borromean cluster sizes can be computed with an adjacency matrix with only $\approx0.2\times L^{3.8}$ components. For larger lattices, the Borromean clustering algorithm becomes computationally unfeasible as the adjacency matrices cannot be accommodated in system memory. 

Our second algorithm for large, dense Borromean lattices iterates through each unvisited node, checking a given node for triangular connections with its neighbors and marking the given node as visited. If neighboring nodes share a Borromean connection, those nodes are idempotently added to a growing list of components connected to the original node. The triangular search procedure is repeated for each node that is added to the list, until no more new connections are found. The algorithm then returns to the initial outer iteration and repeats the connected component search for the next unvisited node. This avoids the need for a large adjacency matrix, and is faster on a laptop than our first algorithm for full lattices above $L\approx50$ and percolating lattices above $L\approx200$.

To perform our computational measurements, we initialized partially occupied square lattices as random binary matrices, varying the fraction of occupied sites, $q$, also known as the site occupation probability. We calculated the cluster size distribution for Borromean and Hopf connectivity rules from size $L=12$ up to $L=1000$. Our algorithms were implemented in MATLAB, and periodic boundary conditions were not used.

The knot-lattice-graph mapping and algorithm in which each node may be Borromean linked in various ways with its eight lattice neighbors represents a fully symmetric square lattice with Borromean connectivity, but \textit{does not} represent the system in Figure 1. In Figure 1, the blue and yellow loops do not form Borromean triplets with each other, whereas the red and green loops do. To describe the connectivity of this network, separate rules are needed for green/red and blue/yellow loops. Green and red loops form Borromean connections with two of their neighbors if one is a nearest neighbor and one is a next-nearest neighbor along the hypotenuse (eight of the twelve possible triangle connections), and yellow and blue loops form Borromean connections with two of their nearest neighbors (four of the twelve possible triangle connections). Results from this mapping are reported in the Appendix.





\section{Results and Discussion}
\subsection{Identifying the Percolation Threshold}
Above the percolation threshold, almost all occupied lattice nodes are within the largest cluster, while below the threshold almost none are. The sharpness of this transition increases with the size of the lattice. The size of the second-largest cluster is maximized at the percolation threshold \cite{margolina1982size}, which is our primary method of determining it from our computations. Figure 3 shows the largest and second largest cluster sizes as a function of the occupation probability for Borromean and Hopf-linked square lattices of width $L=400$. The largest component size varies with occupation probability as expected, with the jump associated with percolation occurring at a higher occupation probability for the Borromean lattice. Examining the second largest component provides more evidence that the Borromean lattice has a higher percolation threshold. We observe that the standard square lattice has a second-largest component that is maximized near the expected value of 0.5927 \cite{ziff1992spanning}. Under Borromean connectivity rules, the percolation threshold is slightly but significantly higher. It is difficult to precisely locate the peak with finite numerical precision and a finite-sized lattice, but we estimate the peak at approximately 0.606. A more precise analysis of a regularly-connected square lattice at $L=400$ finds a second-largest component peak at $q=0.5917$, 0.16\% below the asymptotic value, suggesting that the Borromean percolation threshold is slightly higher than 0.606. Analysis based on the universality of the fractal dimension at percolation (below) puts our estimate at 0.6075.

\begin{figure}
    \centering
    \includegraphics[width=\textwidth]{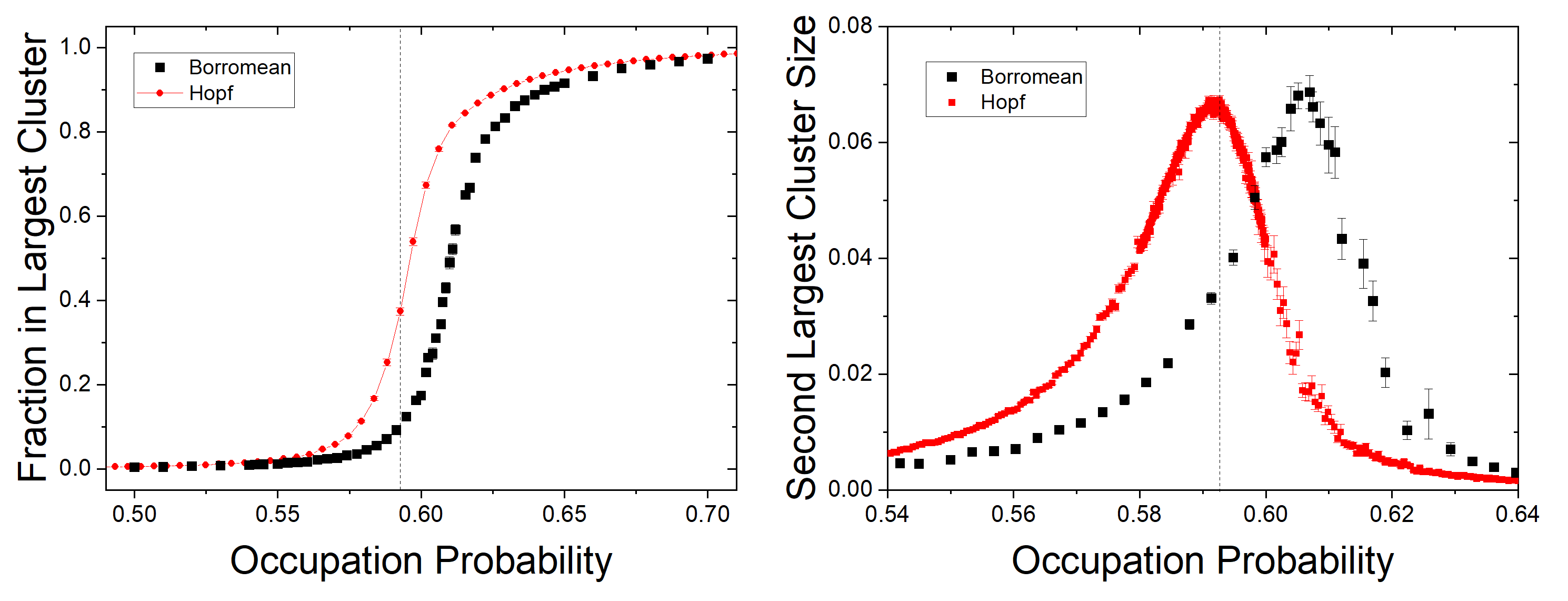}
    \caption{Left: Fraction of occupied sites in a 400$\times$400 square lattice that are within the largest component, as a function of fraction of sites that are occupied, for the Hopf and Borromean connectivity rules.   Right: The second largest component sizes from the same data. The peak of the second largest component indicates the location of the percolation threshold, which is slightly higher for Borromean lattices. In both charts, error bars represent standard error over multiple randomized lattice instances and the established square lattice percolation threshold of 0.5927 is shown by a dashed line.}
    \label{fig:my_label}
\end{figure}

The higher occupation ratio required for percolation with Borromean connectivity is illustrated in Figure 4. The largest cluster at the square lattice percolation threshold will span the entire system (in this case defined as touching the four edges of the lattice), but consist of regions where the cluster forms ``bridges'' only one site wide and several sites long. These can be seen at the blue-teal boundaries in Figure 4. Under the Borromean connectivity rule, the giant cluster does not span these bridges, and forms additional smaller Borromean clusters. Extra occupied sites adjacent to these bridges are required to keep them in the giant cluster, which is why a slightly higher occupation is required for a Borromean network to percolate. Another argument, put forth by Robert Ziff in a personal communication, is as follows: at the square lattice percolation threshold of $q=0.5927$, the probability that two adjacent sites are filled (and form a cluster) is $q^{2}=0.3513$. However, for the sites to form a Borromean triplet, one of the four sites adjacent to the doublet must be occupied, which occurs with probability $q^{2}(1-(1-q)^4)$. For that probability to equal 0.3513, $q$ must have a value of 0.6004, slightly higher than 0.5927 but slightly slower than what we observe.

Results from the modified connectivity rules that describe the network in Figure 1 are presented in the Appendix. We note that the difference between Borromean and Hopf connectivity in that case is much stronger, and the percolation threshold is located at approximately 0.66.
\begin{figure}
    \centering
    \includegraphics[width=0.4\textwidth]{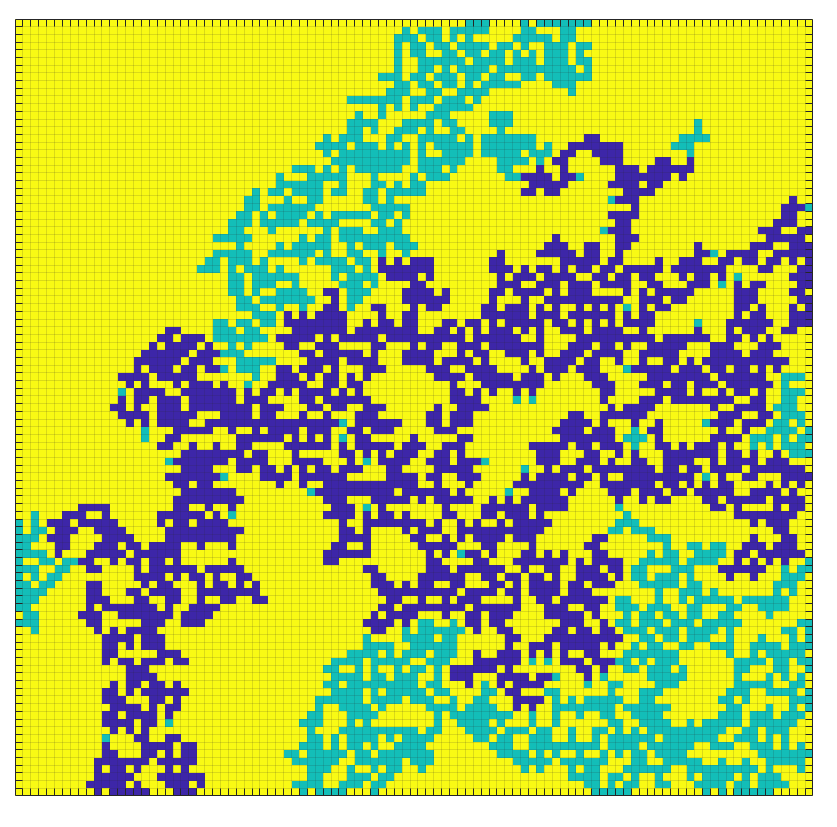}
    \caption{A percolating cluster touching all four edges of a 100$\times$100 square lattice at the percolation threshold of 0.5927. For clarity only the largest cluster is shown, as the union of the teal and blue sites, which touches all four sides of the lattice. The blue subset is the largest component under the Borromean connectivity rule, and this cluster does not span the entire system, only touching two sides. }
    \label{fig:perczample}
\end{figure}
\subsection{Universal Behavior}
Although the connectivity of the Borromean lattice is not pairwise, it is still local, and at percolation is expected to obey the expected universal behavior of percolating systems. In two dimensions, this includes a fractal dimension for largest cluster size of $91/48\approx1.89$ \cite{ziff2011correction}, and a cluster size distribution that decays with a -187/91 $\approx2.05$ power law \cite{macleod1998large}. The largest component size at $q=0.6075$ is plot against the system size in Figure 5, along with data for the Hopf lattice at 0.5927, which is essentially identical. The best fit value for the fractal dimension is 1.893$\pm$0.002, or (90.9$\pm$0.1)/48. If the largest cluster size from systems that are not at the percolation threshold is examined, it is expected to slightly deviate from power-law behavior, and display a best-fit exponent that is not the universal value. At $q=0.606$, we observed an exponent of 1.82$\pm$0.02, and at 0.609 we observed 1.96$\pm$0.02. We interpret this as additional evidence that the percolation threshold of the Borromean square lattice is at approximately 0.6075. 

\begin{figure}
    \centering
    \includegraphics[width=\textwidth]{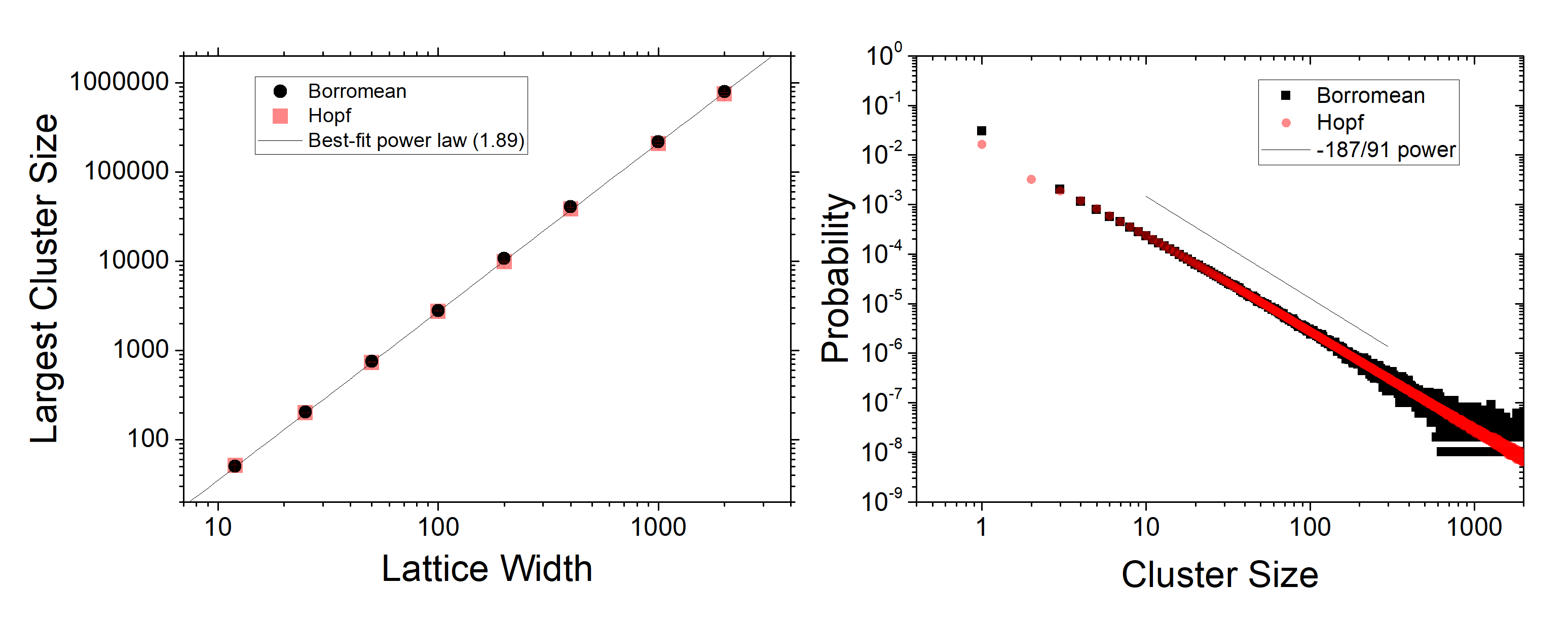}
    \caption{Universal properties of the percolation threshold. Left: The largest component size as a function of the lattice width for Borromean lattices at $q=0.6075$, with a best-fit exponent of 1.893$\pm$0.002, consistent with a fractal dimension of 91/48. Data for regular square connectivity at $q=0.5927$ is overlaid and effectively identical. Error bars are smaller than points. Right: Cluster size probability distribution for 1000$\times$1000 lattices with Borromean and Hopf connectivity at their respective percolation thresholds. Both distributions are consistent with each other, although presumably due to finite-size effects have a weaker decay exponent than the expected -187/91. The Hopf data is more precise due to the faster algorithm used to generate it.}
    \label{fig:universal}
\end{figure}

The cluster size distribution at percolation is seen in Figure \ref{fig:universal} for both Borromean and Hopf connectivity rules. The distribution is slightly different for the smallest clusters, notably due to the lack of Borromean dimers, but they converge to a similar form for large clusters. We measure best-fit exponents for these distributions as -1.883$\pm$0.001 for Borromean and -1.906$\pm$0.001 for Hopf connectivity, inconsistent with the Fisher exponent of -187/91. The cluster size distribution exponent is more sensitive to finite-size effects: Ding et al. observed a power-law consistent with a -1.92 exponent on square lattices 1000 sites wide \cite{ding2014numerical}, while in an earlier study Hoshen et al. observed data consistent with a -1.78 exponent on triangular lattices 4000 sites wide \cite{hoshen1979monte}. Only on lattices 2 million sites wide have precise data consistent with the -2.05 exponent been observed \cite{macleod1998large}. Although finite-size effects persist, we find that Borromean lattices at the percolation threshold display the same universal properties as those with traditional connectivity.
\subsection{Dissolution Spectrum}
An experiment to determine the topology of a hypothetical Borromean material may break down the network in a similar way that kinetoplasts are broken down by enzymes, and examine the topology of the smaller clusters that are released from the network (for example with atomic force microscopy). How might a Borromean material be distinguished from a Hopf-linked catenated network? 

Chen et al. \cite{chen1995topology} discuss the probability of isolated rings as well as Hopf-linked doublets and triplets being released from kinetoplast networks during enzyme digestion, assuming different lattice geometries. The isolation probabilities depend on the linearization probability $p$ of a minicircle being cut by the enzyme, equivalent to a node being removed from a graph, as well as the co-probability $q=1-p$, which is equivalent to our lattice occupation probability. We may calculate similar probabilities for Borromean connectivity on the square lattice. To find the probability of isolating a single ring from the network, we consider a 3x3 lattice with the central node occupied, and find that 34 of the $2^8$ possibilities for the surrounding nodes being occupied lead to the central node being isolated.  Summing the probability of each configuration yields:
\begin{equation}
    P_{1B}=p^{4}q^{5}+8p^{5}q^{4}+16p^{6}q^{3}+8p^{7}q^{2}+p^{8}q.
\end{equation}

To find the probability of a Borromean triplet being released from the network, we consider three occupied sites in an L-shape in the center of a 4x4 grid, and find that 100 of the $2^{13}$ possibilities lead to the links being released. The total probability is:

\begin{equation}
    P_{3B}=4\left(p^{7}q^{9}+8p^{8}q^{8}+24p^{9}q^{7}+24p^{10}q^{6}+24p^{11}q^{5}+8p^{12}q^{4}+p^{13}q^{3}\right),
\end{equation}
the factor of 4 admitting rotations of the triplet. For comparison, the probabilities for finding an isolated component or a triplet with regular square lattice Hopf connectivity are $P_{1H}=qp^4$ and $P_{3H}=2q^{3}p^{7}(2+p)$.

\begin{figure}
    \centering
    \includegraphics[width=\textwidth]{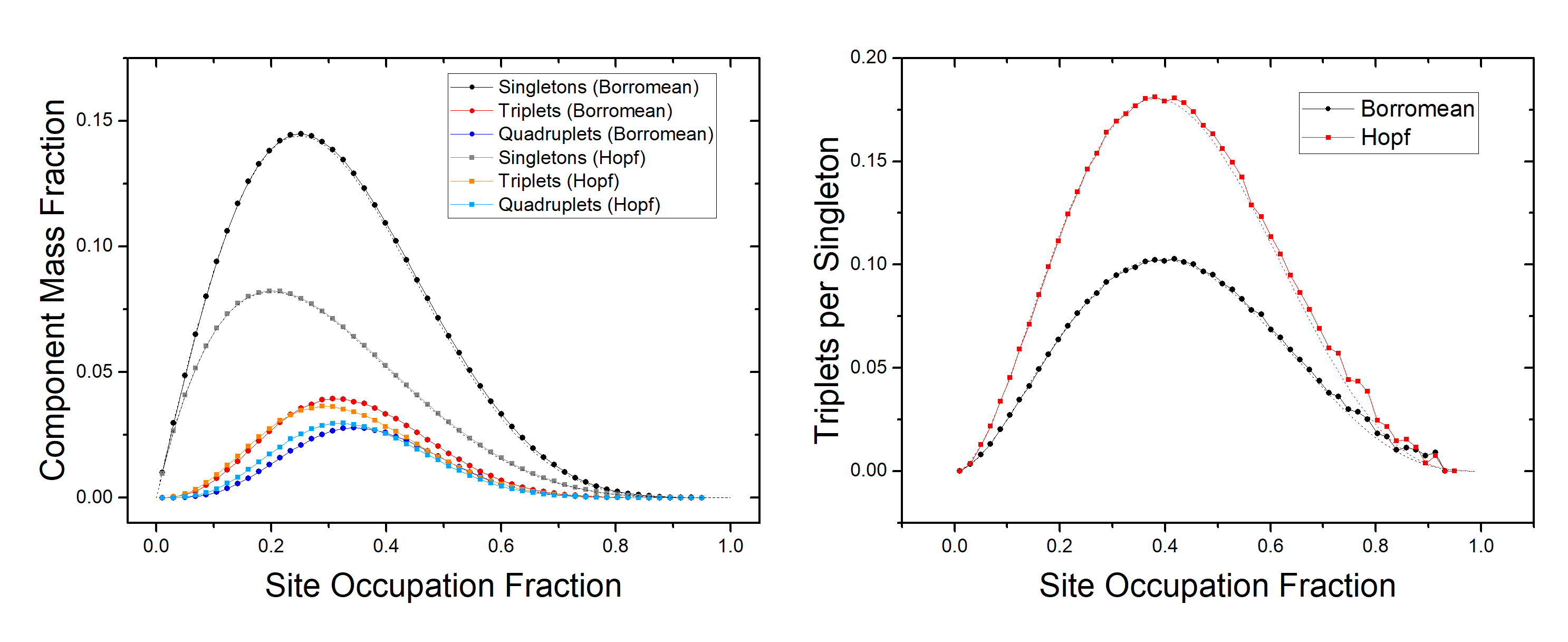}
    \caption{Left. Fraction of total mass found in isolated loops, three-component loops, and four-component loops after a dissolution reaction of Borromean and Hopf-linked topological networks, simulated for 200x200 lattices. The exact predictions for the singlet distribution are overlaid, with slight discrepancies due to edge effects. Right. The ratio of triplets to singlets after a dissolution under Borromean and Hopf connectivity.}
    \label{fig:dissol}
\end{figure}

Figure \ref{fig:dissol} shows the expected mass fraction of different sized components as a function of the undissolved fraction $q$, for Borromean and Hopf networks. A greater proportion of the mass is found in single loops under Borromean connectivity, in part due to the absence of 2-component links which occupy part of the mass in dissolved Hopf systems. There are subtle differences in the relative mass found in 3- and 4-component links between the two connectivity rules. The ratio of triplets to singlets, a parameter that could in principle be measured experimentally (and has been, for kinetoplast DNA), is shown on the right, with the ratio for Borromean networks peaking close to 20\% near $q=0.4$, but only 10\% for Hopf networks. The exact expressions describe the computational data well, with some finite-size edge discrepancies at large $q$.

\section{Conclusion}

We have simulated the dissolution of Borromean square lattices to calculate their percolation threshold, which we find to be 2.4\% higher than that of the traditional square lattice. We have also made predictions about the population of simple links released from a network during dissolution, finding marked differences from Hopf lattices due in part to the lack of Borromean dimers. In addition to being the first such analysis, it also makes predictions for potential experiments on Borromean materials \cite{thorp2015infinite}. 

This analysis is possible because the square lattice structure allows us to map a knot theory problem, the Borromean connection of three loops, to a graph theory problem, the continuity of connected triangles. More general linked-ring systems cannot necessarily rely on geometry to impose topology, but invariants have been derived which can detect triplet-wise Borromean connectivity when pairwise linking connectivity is absent \cite{deturck2009triple}. A number of studies have used the Gauss linking number to calculate the graph structure of densely packed circles \cite{diao2012effects,rodriguez2013percolation} or ring polymers \cite{michieletto2015kinetoplast,d2017linking}, but the linking number cannot detect Borromean connectivity. Building upon these results, future work can examine the Borromean networks that form within dense packings once Hopf connections have been accounted for. This analysis need not end at square lattices, nor at Borromean connectivity: lattices may be constructed of any N-connected Brunnian link, which may present a range of percolation phenomena.

\section{Acknowledgements} The authors wish to thank Robert Ziff for his helpful discussions about percolation theory. They are also grateful to Kenny Umenthum for suggesting a recursive algorithm, Raphael Candelier for his helpful blog, and Luc Devroye whose artwork inspired this project. DF is supported by the HSI-STEM Bridge to the Beach Program. This work is supported by the National Science Foundation, grant number 2105113.

\bibliographystyle{unsrt}
\bibliography{borrorefs.bib}

\section{Appendix}

For completeness, we present results for the second-largest component of Borromean lattices that correspond to Figure 1, in which red and green loops share Borromean links with each other mediated by yellow or blue loops, but blue and yellow loops do not share Borromean links mediated by red or green loops. Compared to the generalized lattice mapping we describe in the main text, this version breaks translational symmetry. To count connectivity between between sites, green and red sites must form a triangle with one nearest neighbor and one next-nearest neighbor, while blue and yellow sites must form a triangle with two nearest neighbors. As can be seen from Figure 7, the deviation from the square lattice is stronger, with the percolation threshold occurring at approximately $q=0.66$.

\begin{figure}
    \centering
    \includegraphics[width=0.75\textwidth]{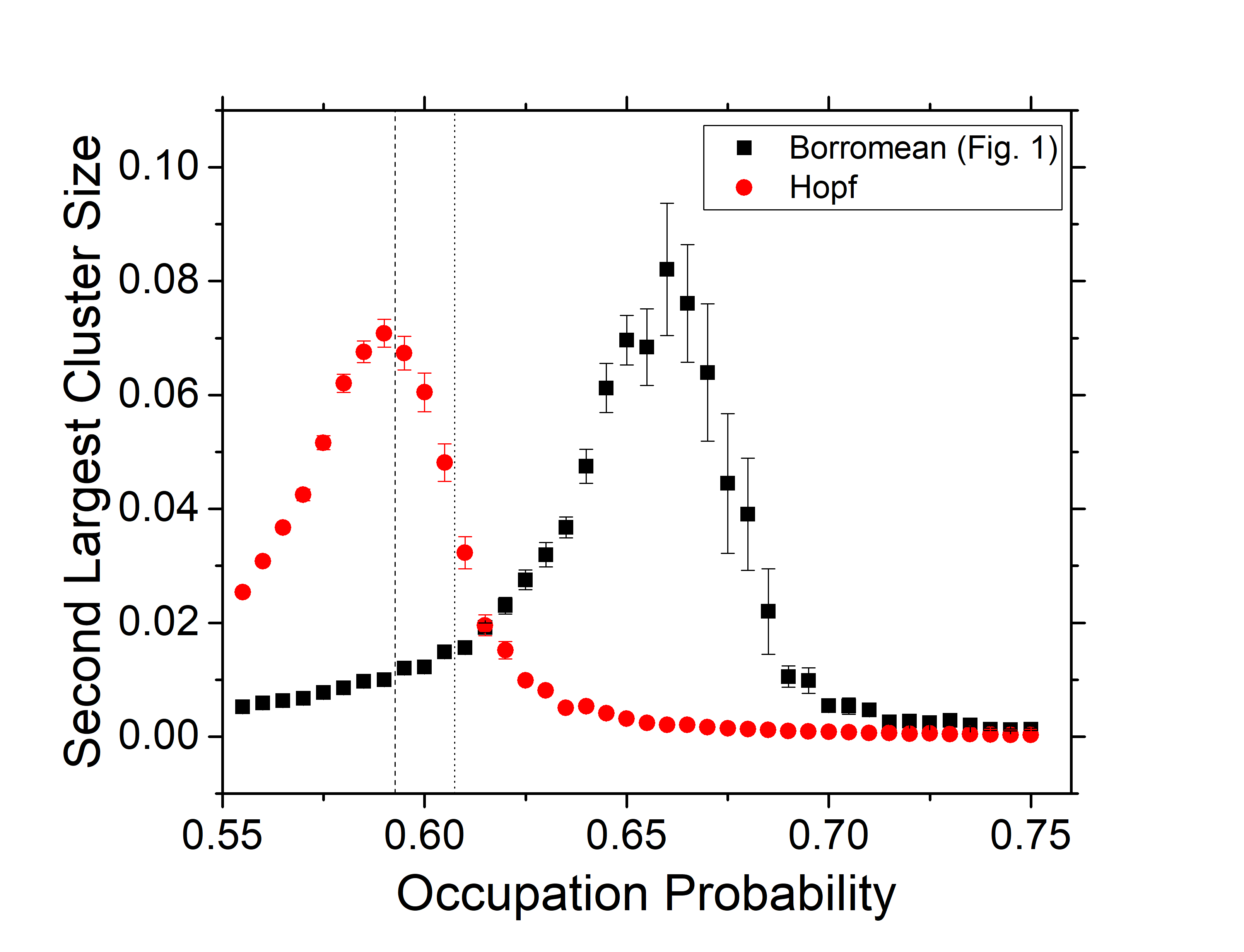}
    \caption{The second largest component size for 200$\times$200 square lattices corresponding to the network in Figure 1. The vertical lines are at 0.5927 and 0.6075, the percolation thresholds for traditional square lattice connectivity, and fully symmetric Borromean square lattice connectivity. The peak is located close to 0.66.}
    \label{fig:appendix}
\end{figure}

\end{document}